\documentstyle[hf97_sprocl]{article}
\input{psfig}
\begin{document}

\title{Light meson and baryon spectroscopy from charm decays in Fermilab E791}

\author{M. V. Purohit\\ for the Fermilab E791 Collaboration}
\address{Department of Physics \& Astronomy, \\
University of South Carolina, Columbia, SC 29208, USA \\
e-mail: purohit@sc.edu
}
\maketitle\abstracts{
      We present results from Fermilab experiment E791. We extracted the
fractions of resonant components in the $\Lambda_c^+\to p K^- \pi^+$
decays, and found a significant polarization of the $\Lambda_c^+$ using a fully
5-dimensional resonant analysis. We also did resonant analyses of $D^+$
and $D^+_s$ decays into $\pi^+\pi^-\pi^+$. We observed an insignificant
asymmetry in the Breit Wigner describing the $f_0(980)$ and found good
evidence for a light and broad scalar resonance in the $D^+$ decays.
}

\section{Experiment E791 at Fermilab}

      Fermilab E791 is a fixed-target pion-production charm 
collaboration involving 17 institutions. Over 20 billion events were 
collected during 1991-92.
The experiment used a segmented target with five thin foils (1 Pt, 4 C)
with about 1.5 cm center-to-center separation. With this target 
configuration and with 23 planes of silicon detectors we were able to suppress 
the large backgrounds due to random track combinations and secondary
interactions. Complementing this vertex detector is a complete 2-magnet 
spectrometer with 35 planes of drift chambers and with Cherenkov detectors and 
calorimeters for particle identification and energy
measurement.\cite{e791_pairs}

\def\Kbar{$\overline{K}$}
\def\ckv{\v Cerenkov\ }
\def\lbrk{\hfil\break}

\section{$\Lambda_c^+\to p K^- \pi^+$ decays}

This analysis\cite{e791_pkpi} is unique in at least two respects: this is the first
coherent amplitude analysis of a charm baryon decay and is the first fully
5-dimensional resonant analysis of any decay.\cite{pdg}
The most obvious components of 
the $\rm{\Lambda_c^+\to pK^-\pi^+}$ decays (and charge conjugate 
decays, which are implied throughout this paper) include the 
nonresonant $\rm{pK^-\pi^+}$ decay, and the $\rm{p\overline{K}^{*0}(890)}$ 
and $\rm{\Lambda(1520)\pi^+}$ two-body decays. These three decays 
can be described by spectator and W-exchange amplitudes. However, in 
lowest order, the $\rm{\Delta^{++}(1232)K^-}$ decay can occur 
only via the exchange amplitude. Thus, a significant presence of
$\rm{\Delta^{++}(1232)K^-}$ decays would indicate that exchange
amplitudes are important. Unlike charm meson decays, helicity and
form-factor suppression are not expected to inhibit 
exchange amplitudes for charm baryons.

The $\Lambda_c^+$ and its decay proton carry spin, and the
$\Lambda_c^+$ may be polarized upon production. Therefore an analysis
of these decays requires five kinematic variables for a complete
description. As a by-product of the analysis, the production
polarization of the $\rm{\Lambda_c^+}$, ${\bf P}_{\rm{\Lambda_c}}$, is
also measured.  

Fitting to 886$\pm$43 events, we find the fractions of resonant
components listed in Table \ref{pkpi_fitfr}, indicating that exchange
amplitudes (signaled by the $\rm{\Delta^{++}(1232)K^-}$ decays)
contribute significantly to charm baryon decays.

\begin{table}[htbp]
\caption{The decay fractions for $\rm{\Lambda_c^+\rightarrow pK^-\pi^+}$
with statistical and systematic errors from the final fit.}
\vspace*{3pt}
\label{pkpi_fitfr}
\begin{tabular}{cc}\hline
Mode & Fit Fraction ($\%$) \\ \hline
\vspace*{-10pt} & \\
$\rm{p\overline{K}^{*0}(890)}$	& 19.5$\pm$2.6$\pm$1.8 \\
$\rm{\Delta^{++}(1232)K^-}$		& 18.0$\pm$2.9$\pm$2.9 \\
$\rm{\Lambda(1520)\pi^+}$	&  7.7$\pm$1.8$\pm$1.1 \\ 
Nonresonant         		& 54.8$\pm$5.5$\pm$3.5 \\ \hline
\end{tabular}
\end{table}

The projections of the data and of the fit on
the traditional $m^2$ variables as well as on the angular variables
introduced by the spin-dependent 5-dimensional analysis show that we get
had a good fit (the $\chi^2$/DF is 1.06). The only
discrepancy between data and the fit is in the low $m_{pK}^2$
region. This may be due to the tail of a $\Lambda(1405)$ decaying to
$pK^-$ (see Ref. \cite{dal98}). We also searched for other resonances and found that the data
weakly favor a $\frac12^-$ resonance in $pK$ with mass 1556 $\pm$ 19 MeV/c$^2$ and 
width 279 $\pm$ 74 MeV/c$^2$. However, no such resonance is known to
exist. 

\begin{figure}[htbp]
\psfig{figure=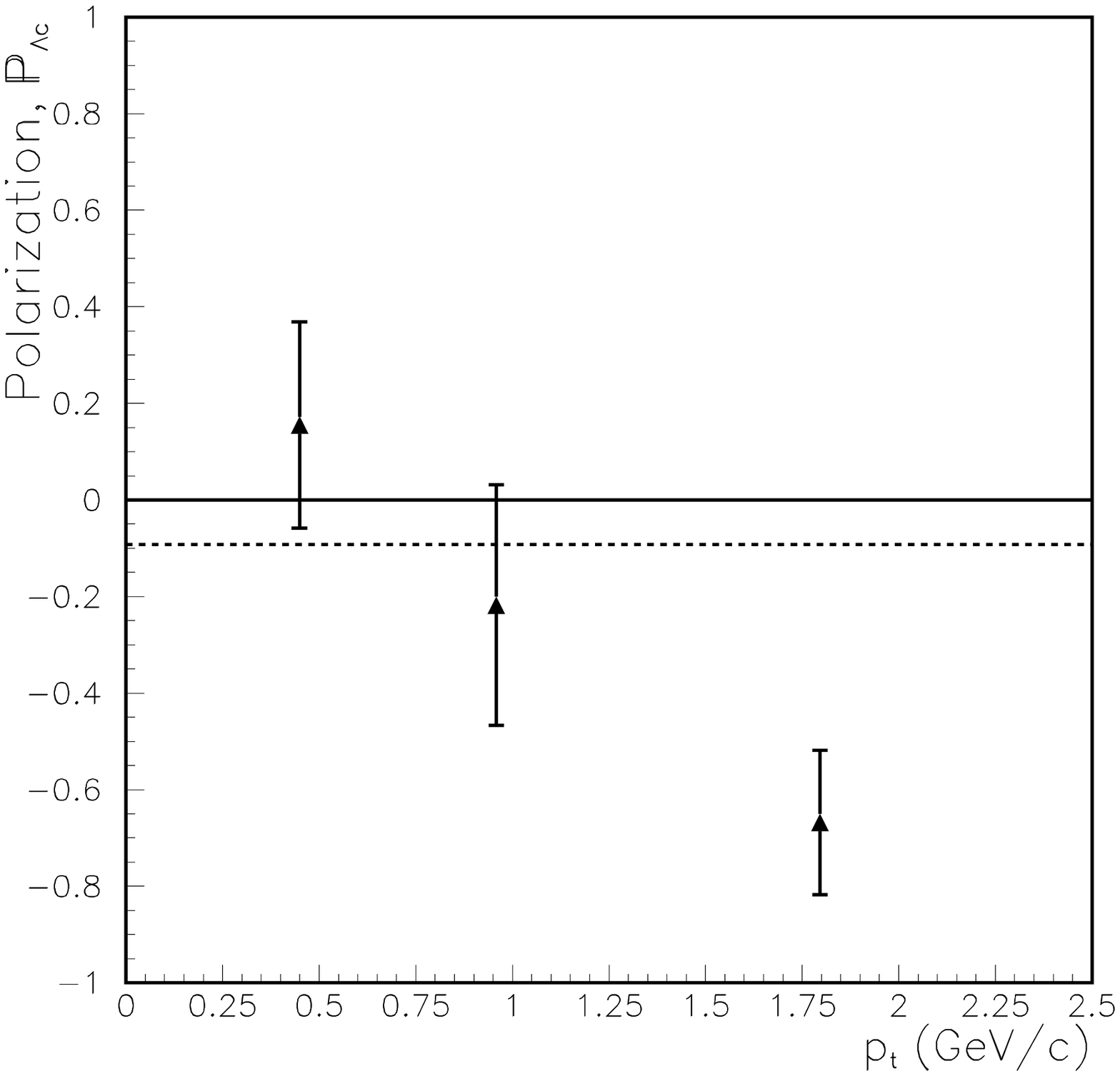,height=2.5in,width=4.0in}
\caption{The polarization of the $\rm{\Lambda_c}$ as a function of the
         $\rm{\Lambda_c}$'s transverse momentum. 
         \label{ppt}
        }
\end{figure}

In Figure \ref{ppt} we show the result of measuring the polarization as
a function of $p_T^2$. The average polarization $P$ for all our data is
consistent with zero but there is a clear fall in $P$ as a function of
$p_T^2$. 

\section{Dalitz analyses of $D^+$, $D^+_s$ decays to $\pi^+\pi^-\pi^+$}

      In Figure \ref{pipipi_mass} below, the $\pi^+\pi^-\pi^+$ mass plot has
1172$\pm$61 events above background in the $D^+$ signal region and
848$\pm$44 events above background in the $D^+_s$ signal region. 

A fit to the $D^+_s$ decays\cite{e791_3pi} to $\pi^+\pi^-\pi^+$ followed the WA76
parameterization for the $f_0(980)$ resonance\cite{wa76} and yields the
parameters $m[f_0(980)] = 977\pm 3\pm 2$ MeV, $g_K = 0.02 \pm 0.04 \pm
0.03$ and $g_\pi = 0.09 \pm 0.01 \pm 0.01$, in disagreement with results
from WA76. The fit fractions are shown in Table \ref{pipipi_fitfr}. Note
that there is no significant fraction of decays into purely non-strange
modes, indicating that the annihilation amplitude is small.

The fit to the $D^+\to \pi^+\pi^-\pi^+$ decays had poor quality until we
tried to include a scalar resonance. We found that including such a
resonance improves the fit quality from a $\chi^2$/DF of 254/162 (poor)
to an acceptable $\chi^2$/DF of 138/162. When the mass and width of this
scalar (the $\sigma$) are allowed to float, we obtain $m_\sigma =
478^{+24}_{-23}\pm 17$ MeV/c$^2$ and $\Gamma_\sigma = 324^{+42}_{-40}\pm
21$ MeV/c$^2$. As additional tests, we fit for a new resonance with a
real scalar amplitude, or as a vector or a tensor particle. In all
three cases, the fit was poorer than the fit with a ``normal'' scalar
resonance. We conclude that we have good evidence for this scalar
resonance and a good physical environment for measuring its mass and
width.

\begin{table}[htbp]
\caption{The decay fractions for $D^+$, $D^+_s$ decays to $\pi^+\pi^-\pi^+$}
\vspace*{3pt}
\label{pipipi_fitfr}
\begin{tabular}{ccc}\hline
\vspace*{-10pt} & \\
Mode                & $D^+_s$ Fit Fraction ($\%$) & $D^+$ Fit Fraction ($\%$) \\ \hline
Non-resonant        &  0.5$\pm$1.4$\pm$1.7 &  7.8$\pm$6.0$\pm$2.7 \\
$f_0(980)\pi^+$     & 56.5$\pm$4.3$\pm$4.7 &  6.2$\pm$1.3$\pm$0.4 \\
$\rho^0(770)\pi^+$  &  5.8$\pm$2.3$\pm$3.7 & 33.6$\pm$3.2$\pm$2.2 \\
$f_2(1270)\pi^+$    & 19.7$\pm$3.3$\pm$0.6 & 19.4$\pm$2.5$\pm$0.4 \\
$f_0(1370)\pi^+$    & 32.4$\pm$7.7$\pm$1.9 &  2.3$\pm$1.5$\pm$0.8 \\
$\rho^0(1450)\pi^+$ &  4.4$\pm$2.1$\pm$0.2 &  0.7$\pm$0.7$\pm$0.3 \\
$\sigma\pi^+$       &  \omit               & 46.3$\pm$9.0$\pm$2.1 \\ \hline
\end{tabular}
\end{table}

\begin{figure}[htbp]
\psfig{figure=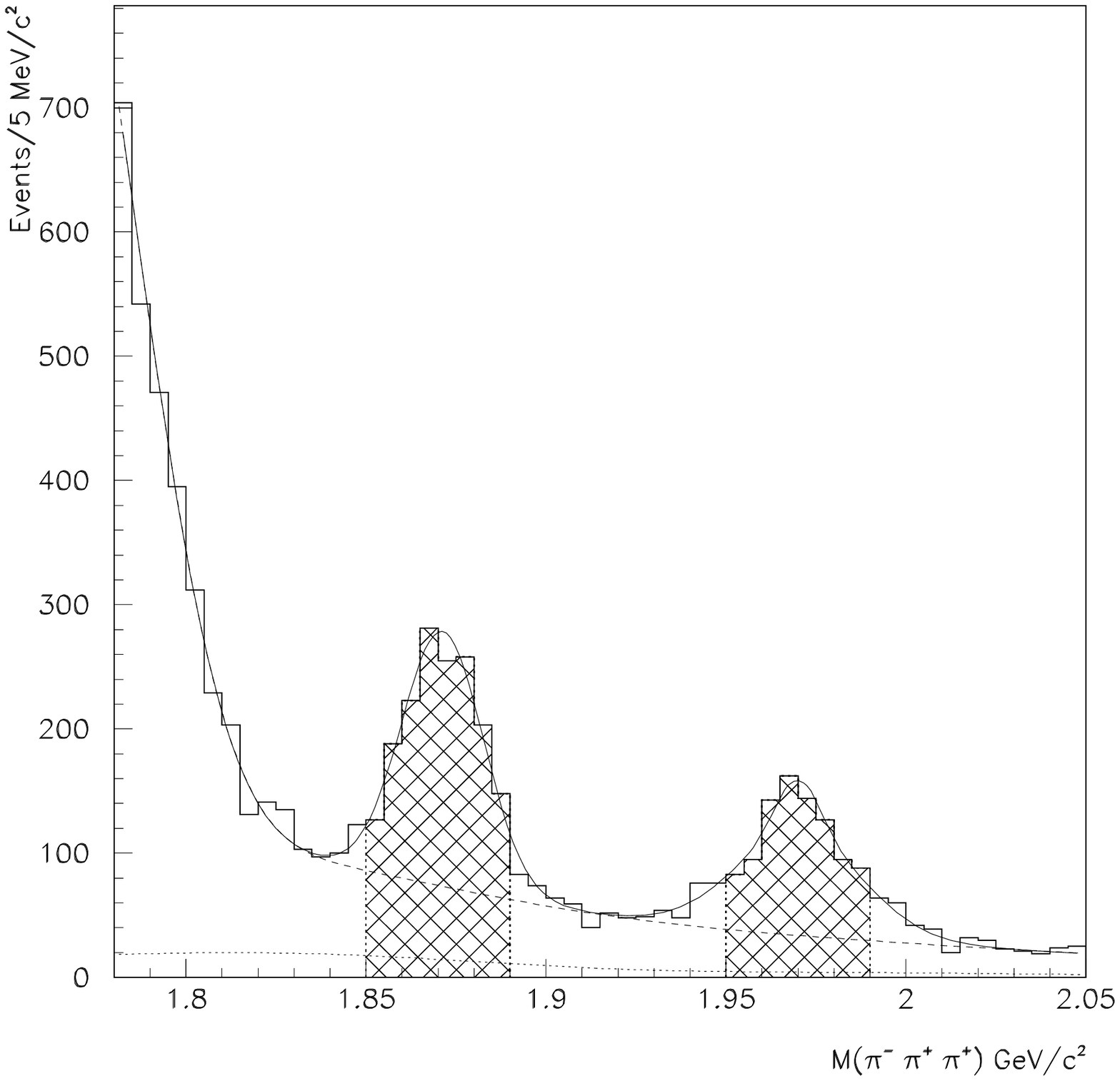,height=2.5in,width=4.0in}
\caption{The three pion mass showing the $D^+$ and $D^+_s$ peaks used
         for the Dalitz analysis
         \label{pipipi_mass}
        }
\end{figure}

\section{Acknowledgements}

      I would like to thank members of my collaboration (Fermilab
E791). This work was supported by a grant from the U.S. Department of
Energy.

\end{document}